\pdfoutput=1

\documentclass[prl,twocolumn,showpacs,preprintnumbers,amsmath,amssymb,
superscriptaddress,nofootinbib]{revtex4-1}
\usepackage{epsfig}
\usepackage{graphicx}
\usepackage{color}
\usepackage[unicode=true,pdfusetitle,
bookmarks=true,bookmarksnumbered=false,bookmarksopen=false,
breaklinks=false,pdfborder={0 0 1},backref=false,colorlinks=false]
{hyperref}

\newcommand{\nn}{\nonumber}
\newcommand{\be}{\begin{equation}}
\newcommand{\ee}{\end{equation}}
\newcommand{\bea}{\begin{eqnarray}}
\newcommand{\eea}{\end{eqnarray}}
\newcommand{\Eq}[1]{Eq.~(\ref{#1})}

\begin{document}

\preprint{\hbox{CALT-2015-047} }

\title{On-Shell Recursion Relations for Effective Field Theories}

\author{Clifford Cheung}
\affiliation{Walter Burke Institute for Theoretical Physics, California
Institute of Technology, Pasadena, CA, USA}
\author{Karol Kampf}
\affiliation{Institute of Particle and Nuclear Physics, Charles University, Prague, Czech Republic}
\author{Jiri Novotny}
\affiliation{Institute of Particle and Nuclear Physics, Charles University, Prague, Czech Republic}
\author{Chia-Hsien Shen}
\affiliation{Walter Burke Institute for Theoretical Physics, California
Institute of Technology, Pasadena, CA, USA}
\author{Jaroslav Trnka}
\affiliation{Walter Burke Institute for Theoretical Physics, California
Institute of Technology, Pasadena, CA, USA}

\date{\today}

\begin{abstract}

We derive the first ever on-shell recursion relations for amplitudes in effective field theories.  Based solely on factorization and the soft behavior of amplitudes, these recursion relations employ a new rescaling momentum shift to construct all tree-level scattering amplitudes in theories like the non-linear sigma model, Dirac-Born-Infeld theory, and the Galileon.  Our results prove that all theories with enhanced soft behavior are on-shell constructible.

\end{abstract}

\maketitle

\section{Introduction}

The modern S-matrix program exploits physical criteria like Lorentz invariance and unitarity to construct scattering amplitudes directly and without the aid of a Lagrangian.  
At tree level, many S-matrices are constructible via on-shell recursion, which elegantly encodes factorization as a physical input.  Originally discovered in the context of gauge theory \cite{BCFW}, on-shell recursion relations were soon extended to gravity theories \cite{GR} and, eventually, all renormalizable and some non-renormalizable theories \cite{Recursion, OurRec}.  Subsequently, these developments led to progress in alternative formulations of quantum field theory, {\it e.g.}, in the context of planar ${\cal N}=4$ super Yang-Mills theory through the positive Grassmannian~\cite{Grassmanian} and Amplituhedron~\cite{amplitudhedron}.  There has, however, been remarkably little progress towards a fully on-shell formulation of effective field theories.  Such an omission is unfortunate, as effective field theories provide a universal description of spontaneous symmetry breaking in all branches of physics, ranging from superconductivity to the strong interactions~\cite{pion} to cosmology~\cite{Cheung:2007st}.

The aim of this letter is to fill this gap.  We derive a new class of recursion relations that fully construct the S-matrices of scalar effective field theories by harnessing an additional physical ingredient: the vanishing of amplitudes in the soft limit.  This approach is logical because the soft behavior of the S-matrix actually dictates the interactions and symmetries of the corresponding effective field theory \cite{OurEFT}, thus giving a theory classification purely in terms of on-shell data.  Our new recursion relations apply to any theory with enhanced soft limits, including the non-linear sigma model, Dirac-Born-Infeld theory, and the Galileon~\cite{Galileons}.

\section{Recursion and Factorization}

On-shell recursion relations act on an initial seed of lower-point on-shell amplitudes to bootstrap to higher-point.  Criteria like Lorentz invariance---which prescribes strict little group covariance properties of the amplitude \cite{ElvangHuang}---are manifest provided the initial amplitudes and recursion relation maintain these properties at each step.  

The property of factorization, on the other hand, enters less trivially.  
To access multiple factorization channels, the BCFW recursion relations \cite{BCFW} employ a complex deformation of two external momenta,
\begin{equation}
p_1 \rightarrow  p_1+zq\quad \textrm{and} \quad p_2 \rightarrow p_2 - zq,
\label{BCFWshift}
\end{equation}
where $q$ is fully fixed  up to rescaling the on-shell conditions $q^2= q\cdot p_1=q  \cdot p_2=0$. The original amplitude is extracted from the complexified amplitude $A_n(z)$ by contour integrating over an infinitesimal circle centered around $z=0$. Cauchy's theorem then yields a new expression for the original amplitude,
\begin{equation}
A_n(0) = \frac{1}{2\pi{ i}}\oint \frac{dz}{z} A_n(z)=- \sum_I \underset{ z = z_I }{\rm Res}\left(\frac{A_n(z)}{z}\right),
\label{BCFW}
\end{equation}
where $I$ labels factorization channels at which the intermediate momentum $P_I(z)$ goes on-shell, so $z_I$ is defined by $P_I(z_I)^2=0$.  The residue at each pole is 
\bea
-\underset{ z= z_I}{\rm Res}\left(\frac{A_n(z)}{z}\right) &=& A_{n_I}(z_I)\frac{1}{P_I^2}A_{\bar n_{I}}(z_I)\label{prod},
\eea
establishing a recursion relation in terms of the lower-point amplitudes $A_{n_I}$ and $A_{\bar n_{I}}$ where $n_I+\bar n_{I}= n+2$.

The above derivation fails when there is a non-zero residue at $z=\infty$.  However, this boundary contribution is calculable in certain circumstances \cite{Boundary} and moreover there exist any number of generalizations of BCFW recursion for which the amplitude vanishes at large $z$ \cite{Recursion, OurRec}.   Ultimately, this is not surprising because the boundary term literally encodes a class of factorization channels \cite{NimaLargeZ}.
Since BCFW recursion and its extensions apply to all renormalizable and some non-renormalizable theories \cite{Recursion, OurRec}, the corresponding S-matrices are completely fixed by Lorentz invariance and factorization.

\section{Recursion and Soft Limits}

In effective field theories, BCFW recursion and its generalizations are hindered by a non-zero boundary term at $z=\infty$.\footnote{In previous work \cite{NLSMlet}, we derived semi-off-shell recursion relations for the non-linear sigma model, though these methods do not generalize straightforwardly.}  Naively, this is attributable to the divergent behavior of non-renormalizable interactions at large momenta, but this is plainly false in gravity theories, which have terrible high energy behavior but are perfectly constructible via BCFW recursion.  For effective field theories, the problem is simply more fundamental: amplitudes are {\it not} just fixed by factorization, and additional information is needed. In hindsight this is obvious since high-order contact operators in effective field theories are typically related to low-order contact operators not by factorization but by symmetries.

Since existing recursive technology already exploits amplitudes' singularities, a natural candidate for new physical information is amplitudes' zeros.  The former are dictated by factorization while the latter require special kinematics at which the amplitude vanishes.  Amplitudes in effective field theories typically vanish in the limit that $p\rightarrow 0$ for the momentum of an external particle, so there exists a classification of theories according to the degree of their soft behavior \cite{OurEFT}, $\sigma$, where
\begin{equation}
A_n \sim p^\sigma\quad \textrm{for}\quad p\rightarrow0\label{soft0},
\end{equation}
and $\sigma \geq 1$ is an integer.
As shown in \cite{OurEFT}, higher values of $\sigma$ correspond to more symmetry in the theory.

To exploit \Eq{soft0} we need a momentum shift that probes the soft limits of external legs.  This is not accomplished by the BCFW shift in \Eq{BCFWshift}, which probes collinear but not soft behavior.  For our purposes we define a ``rescaling shift" on all external legs, 
\begin{equation}
p_i\rightarrow p_i(1-za_i),
\label{rescaleshift}
\end{equation}
where the $a_i$ are defined up to an overall rescaling and 
\bea\sum_{i=1}^n a_i p_i = 0,
\label{eq:momcons}
\eea
to maintain momentum conservation.  For $n<D+1$, a generic set of momenta $p_i$ are linearly independent, so the only solution to \Eq{eq:momcons} has all $a_i$ equal, corresponding to total momentum conservation.  Since this momentum shift simply rescales of all the momenta, it is not useful for recursion.  For $n\geq D+1$, \Eq{eq:momcons} is solved by
\bea
a_1 = |\widehat{p_1}p_2 p_3 \cdots p_{D+1}|, \;
a_2 = |p_1\widehat{p_2}p_3 \cdots  p_{D+1}|, \; \ldots \;
\label{Dshift}
\eea
for $a_1,a_2,\ldots, a_{D+1}$ with all other $a_i=0$.
Here the vertical brackets denote determinants whose columns are momenta with the hatted column omitted.  When $n=D+1$, this solution again trivializes to $a_i$ all equal, but for $n > D+1$ it is always possible to find distinct $a_i$ provided $p_i$ represent a general kinematic configuration.

The scaling shift in \Eq{rescaleshift} is  purposely chosen so that 
\begin{equation}
A_n(z) \sim (1-za_i)^\sigma \quad \textrm{for} \quad {z\rightarrow 1/a_i} \label{soft},
\end{equation}
due to \Eq{soft0}, thus recasting the soft behavior as a degree $\sigma$ zero of the amplitude.  To compute the amplitude we apply Cauchy's theorem to a contour encircling all poles at finite $z$
\begin{equation}
\oint \frac{dz}{z}\frac{A_n(z)}{F_n(z)} = 0\label{Cauchy},
\end{equation}
where the denominator factor is defined to be
\bea
F_n(z) = \prod_{i=1}^n (1-a_i z)^{\sigma}.
\eea
The integrand of \Eq{Cauchy} is engineered to be non-singular at $z=1/a_i$ since the poles introduced by $F_n(z)$ are cancelled by zeroes of the amplitude.   Thus, the integrand of \Eq{Cauchy} has poles from factorization channels only, so in analogy with BCFW, the amplitude is
\begin{equation}
A_n(0) = - \sum_I \underset{ z= z_{I\pm}}{\rm Res}\left(\frac{A_n(z)}{z F_n(z)}\right)\label{softRec},
\end{equation}
where $I$ again labels factorization channels.  In contrast with BCFW, the each factorization channel in $P_I(z)$ yields two poles $z_{I\pm}$ corresponding to the roots of
\begin{equation}
P_I^2 + 2 P_I \cdot Q_I z + Q_I^2 z^2 = 0,
\label{poles}
\end{equation}
where $P_I(z) = P_I + z Q_i$ and where 
\bea
P_I = \sum_{i\in I} p_i \quad\textrm{and}\quad Q_I=-\sum_{i\in I}a_i p_i.
\label{PQdef}
\eea  
Each residue is a product of lower-point amplitudes which can be rearranged into a new recursion relation,
\begin{equation}
A_n(0)= \sum_I \frac{1}{P_I^2}  \frac{A_{n_I}(z_{I-})\,A_{\bar n_{ I}}(z_{I-})}{(1-z_{I-}/z_{I+})F(z_{I-})} + (z_{I+} \leftrightarrow z_{I-}).
\label{prod2}
\end{equation}
Again, we assume a vanishing boundary term at $z=\infty$, which is achievable because $F_n(z)$ substantially improves the large $z$ behavior of the integrand of \Eq{Cauchy}.  In the next section we determine the precise conditions under which the boundary term is zero.

\section{Criteria for On-Shell Constructibility}

Next, we determine the conditions under which the boundary term vanishes and the new recursion relation in \Eq{prod2} applies.  Under the rescaling shift in \Eq{rescaleshift}, all momenta scale as $z$ at large $z$.  Consequently, if the $n$-point amplitude scales with $m$ powers of momenta, then  $A_n(z) \sim z^m$ and $F_n(z) \sim n\sigma$ so
\bea
\frac{A_n(z)}{F_n(z) } \sim z^{m-n\sigma}.
\eea
Demanding falloff at $z=\infty$ implies that
\bea
\begin{array}{c}
\textrm{on-shell} \\
\textrm{constructible}
\end{array} \quad\leftrightarrow \quad m/n < \sigma.
\label{constructiblemn}
\eea
At the level of the contact terms this  is {\it exactly} the condition that soft limit of the amplitude is enhanced beyond the naive expectation given by the number of derivatives per field.   So the set of amplitudes with special soft behavior are on-shell constructible.

To lift the criterion for on-shell constructibility from amplitudes to theories, we adopt the $(\rho,\sigma)$ classification of scalar effective field theories presented in \cite{OurEFT}.  In particular, for operators of the form $\partial^m \phi^n$, we define a derivative power counting parameter 
\begin{equation}
\rho = \frac{m-2}{n-2},
\label{rhodef}
\end{equation}
so that an amplitude of a given $\rho$ can factorize into two lower-point amplitudes of the same $\rho$.  The simplest effective theories have a fixed value of $\rho$ but mixed $\rho$ theories also exist.  The derivative power counting parameter $\rho$  in \Eq{rhodef}, together with the soft limit degree $\sigma$ defined in \Eq{soft0} define a two parameter classification of scalar effective field theories.

In terms of the $(\rho,\sigma)$ classification, the criterion of on-shell constructibility in \Eq{constructiblemn} becomes
\bea
(\rho -1) < (\sigma -1)\left(\frac{1}{1-2/n}\right).
\label{rhosigmabound}
\eea
For an effective field theory to be on-shell constructible requires that recursion relations apply for arbitrarily high $n$.  In the large $n$ limit, \Eq{rhosigmabound} yields a simple condition for on-shell constructibility,
\bea
\begin{array}{c}
\textrm{on-shell} \\
\textrm{constructible}
\end{array} \quad\leftrightarrow \quad \rho \leq \sigma \textrm{ and } (\rho,\sigma)\neq(1,1),
\label{criterion}
\eea
which precisely coincides with the class of theories that exhibit enhanced soft behavior. 

Examples of on-shell constructible theories are the non-linear sigma model $(\rho,\sigma)=(0,1)$, Dirac-Born-Infeld theory $(\rho,\sigma)=(1,2)$, and the general/special Galileon $(\rho,\sigma)=(2,2)/(2,3)$~\cite{OurEFT}\footnote{Theories with higher shift symmetries \cite{shiftsym} violate this bound.}.
Among these theories, we dub those with especially good soft behavior, $\rho = \sigma -1$, ``exceptional'' theories. Exceptional theories have a very interesting property: their soft behavior is not manifest term by term in the Feynman diagram expansion, and is only achieved after summing all terms into the amplitude.  Note the close analogy with gauge invariance in Yang-Mills theory or diffeomorphism invariance in gravity, which similarly impose constraints among contact operators of different valency.  The exceptional theories also play a prominent role in the scattering equations \cite{scatteq} and ambitwistor string theories \cite{ambit}, suggesting a deeper connection between these approaches and recursion.

For the exceptional theories, \Eq{criterion} is more than satisfied, yielding better large $z$ falloff than is even needed for constructibility. 
Thus, our recursion relations generate so-called bonus relations defining identities among amplitudes. In principle this can be exploited, for example by introducing factors of $P_i(z)^2$ into the numerator of the recursion relation to eliminate certain factorization channels from the recursion relation.  This is an interesting possibility we leave for future work.


Finally, let us address a slight caveat to the $z$ scaling arguments discussed above.  While all momenta scale as $z$ at large $z$, it is {\it a priori} possible that cancellations modify the naive scaling of $A_n \sim z^m$ for an amplitude with $m$ derivatives.   This is conceivable because the $a_i$ parameters in the momentum shift are implicitly related by the momentum conservation condition in \Eq{eq:momcons}.  In particular, our recursions would fail if there were cancellations in propagator denominators such that they scaled less severely than $z^2$.
That there is always a choice of $a_i$ for which no such cancellations arise can be shown via proof by contradiction.  In particular, assuming no such choice exists implies that cancellations occur for all values of $a_i$.  But we can always perturb a given choice of $a_i$ away from such a cancellation point by applying an additional infinitesimal momentum shift on a subset of $D+1$ external legs as defined in \Eq{Dshift}.  Thus the starting assumption is false and there are generic values of $a_i$ for which $A_n \sim z^m$ scales as expected.

\section{Example Calculations}

In this section we apply our recursion relations to scattering amplitudes in various effective field theories.   We begin with amplitudes in exceptional theories.  Curiously, the six-point amplitudes in the non-linear sigma model, Dirac-Born-Infeld, and the special Galileon, are, term by term, the ``square" and ``cube" of each other, reminiscent of the result of~\cite{scatteq}. Afterwards, we consider the general Galileon, which is marginally constructible.

\medskip
\noindent {\bf Non-Linear Sigma Model: $(\rho,\sigma)=(0,1)$}
\medskip

As shown in \cite{NLSMpap}, flavor-ordered scattering amplitudes in the non-linear sigma model vanish in the soft limit.  We derive the flavor-ordered six-point amplitude $A_6$ by recursing the flavor-ordered four-point amplitude,
\begin{equation}
A_4=s_{12}+s_{23}.
\label{eq:NLSM_4pt}
\end{equation}
Since $A_6$ has three factorization channels, the recursion relation in \Eq{prod2} takes the form
\begin{equation}
	A_{6}= A_{6}^{(123)}+A_{6}^{(234)}+A_{6}^{(345)},
\end{equation}
corresponding to when $P_{123}$, $P_{234}$, and $P_{345}$ go on-shell.  Consider first the pole at $P^2_{123}(z) = 0$, whose roots are
\begin{equation}
z_{\pm} = \frac{-(P_{123}\cdot Q_{123}) \pm \sqrt{(P_{123}\cdot Q_{123})^2-P_{123}^2Q_{123}^2}}{Q_{123}^2}.
\label{NLSMroots}
\end{equation}
Plugging \Eq{PQdef} into  \Eq{prod2} we obtain
\begin{align}
		A_{6}^{(123)} &= &\frac{B}{P_{123}^2}\;\times \hspace*{-.25cm}  \sum_{\substack{ij\in \{12,23\} \\ kl\in \{45,56\}}}
				C_{ijkl}
+(z_{+}\leftrightarrow z_{-}) ,
\label{NLSMterm}
\end{align}
where for later convenience we have defined
\bea
C_{ijkl} = \frac{s_{ij} s_{kl}}{\prod\limits_{m\notin \{i,j,k,l \}}(1-a_m z_-)},
\eea 
and $B=(1-z_-/z_+)^{-1}$.  We observe that $A^{(123)}_{6}$ is equal to the residue of a new function
\begin{align}
	A_{6}^{(123)} &=  -\underset{z= z_{\pm}}{\rm Res}\left [
	\frac{(s_{12}(z)+s_{23}(z))(s_{45}(z)+s_{56}(z))}{z P^2_{123}(z)F_6(z)} \right ] \nonumber \\
	&= \frac{(s_{12}+s_{23})(s_{45}+s_{56})}{P^2_{123}} \\
	& \quad +\sum_{i=1}^6 \underset{z= z_i}{\rm Res}\left [
	\frac{(s_{12}(z)+s_{23}(z))(s_{45}(z)+s_{56}(z))}{z P^2_{123}(z)F_6(z)}
	\right ],
	 \nonumber
	\label{eq:NLSM_simp}
\end{align}
which we have recast in terms of residues at $z=0$ and $z_i=1/{a_i}$ by Cauchy's theorem.
Summing over factorization channels, we simplify the $z_i=1/{a_i}$
residues to
\begin{equation}
	\sum_{i=1}^6 \underset{z= z_i}{\rm Res}\frac{s_{12}(z)+...}{z F_6(z)}=-(s_{12}+...),
	\label{eq:NLSM_simp2}
\end{equation}
where ellipses denote cyclic permutations and we have again applied Cauchy's theorem. Our final answer is
\begin{equation}
A_6 = \left[\frac{(s_{12}+s_{23})(s_{45}+s_{56})}{P_{123}^2} + \dots\right]  - (s_{12}+\dots),
\end{equation}
which is the expression from Feynman diagrams.

\medskip
\noindent {\bf Dirac-Born-Infeld Theory: $(\rho,\sigma)=(1,2)$}
\medskip

Amplitudes in Dirac-Born-Infeld theory are computed similarly with the notable exception that there is no flavor-ordering, so all expressions are permutation invariant.  The four-point amplitude takes the form
\bea
A_4 = s_{12}^2 +s_{23}^2 + s_{13}^2,
\eea
which is the ``square" of \Eq{eq:NLSM_4pt}.
  The six-point scattering amplitude takes the form
\bea
A_6 = A_6^{(123)} + \ldots,
\label{Aperm}
\eea
where the ellipses denote permutations, totaling to the ten factorization channels of the six-point amplitude.
As in \Eq{NLSMroots}, each factorization channel has two roots in $z$, so recursion yields
\begin{align}
		A_{6}^{(123)} &= &\frac{B}{P_{123}^2}\;\times \hspace*{-.25cm}   \sum_{\substack{i,j\in \{1,2,3\} \\ k,l\in \{4,5,6\}}}
				C_{ijkl}^2
+(z_{+}\leftrightarrow z_{-}) ,
\label{DBIterm}
\end{align}
which like before can be shown to be equal to the Feynman diagram expression.  Interestingly, \Eq{DBIterm} is precisely the ``square" form of \Eq{NLSMterm}.

\medskip
\noindent {\bf Special Galileon: $(\rho,\sigma)=(2,3)$}
\medskip

Next, consider the special Galileon \cite{OurEFT,scatteq}, whose existence was conjectured in \cite{OurEFT} due to the existence of an S-matrix with the same derivative power counting as those in the Galileon but with even more enhanced soft behavior (at the same time the amplitudes in this theory were obtained using scattering equations \cite{scatteq}). Shortly after this work it was shown in \cite{hidden} that this theory is a subset of the Galileon theories with a higher degree shift symmetry related by an $S$-matrix preserving duality \cite{Dualities, DualitiesKN}.

Since the Galileon does not carry flavor, its amplitudes are permutation invariant.  The four-point amplitude is
\bea
A_4 = s_{12}^3 +s_{23}^3 + s_{13}^3,
\eea
which is the ``cube" of \Eq{eq:NLSM_4pt}.  Permutation symmetry implies that the amplitude is again of the form of \Eq{Aperm}, except here we find
\begin{align}
		A_{6}^{(123)} &= &\frac{B}{P_{123}^2}\;\times \hspace*{-.25cm}   \sum_{\substack{i,j\in \{1,2,3\} \\ k,l\in \{4,5,6\}}}
				C_{ijkl}^3
+(z_{+}\leftrightarrow z_{-}) ,
\label{SGALterm}
\end{align}
which is the ``cube" of \Eq{NLSMterm}.

\medskip
\noindent {\bf General Galileon: $(\rho,\sigma)=(2,2)$}
\medskip

Finally, let us compute amplitudes in the general Galileon.  As shown in \cite{DualitiesKN}, each $n$-point vertex of the $D$-dimensional Galileon is a Gram determinant,
\bea
V_n &=& G(\widehat p_1, p_2, \ldots, p_{n}) = G( p_1, \widehat p_2, \ldots, p_{n}) = \ldots,\quad
\label{galvertex}
\eea
which is simply the determinant of the matrix $s_{ij}$ with the row and column corresponding to the hatted momentum removed.  The Gram determinant is by construction symmetric in its arguments.  Furthermore,
\bea
G(\lambda p_1, p_2, \ldots, p_n) &= & \lambda^2 G(p_1,p_2, \ldots,p_n),
\label{homog}
\eea
so crucially, the rescaling shift in \Eq{rescaleshift} acts homogenously on the vertex.  This allows for a major simplification of our recursion relation.  Here we define the seed amplitudes for the recursion to be lower-point amplitudes for $n=4,5,\ldots, D+1$.  

For a concrete example, we now apply our new recursion relations to the eight-point amplitude $A_8$ for the Galileon with just a five-point vertex in $D=4$.  The amplitude factorizes into two five-point amplitudes which are simply vertices, {\it e.g.}, $A_5=V_5=G(p_1,p_2,p_3,p_4)$ and $A_{\bar 5} =V_{\bar 5} =G(p_5,p_6,p_7,p_8)$, with the intermediate leg corresponding to the missing column in the Gram determinant.  We find that 
\bea
\frac{A_{5}(z)A_{\bar 5}(z)}{F_8(z)} = \frac{V_{5}(z)V_{\bar 5}(z)}{F_8(z)} = V_{5}(0)V_{\bar 5}(0),
\eea
where we have applied the homogeneity property from \Eq{homog}, thus canceling factors of $(1-a_i z)^2$ in the numerator and denominator.  Summing over factorization channels, we obtain
\begin{align}
A_8&= \frac{B}{P_{1234}^2}  \frac{V_{5}(0)\,V_{\bar 5}(0)}{(1-z_{I-}/z_{I+})} + (z_{I+} \leftrightarrow z_{I-}) +\ldots \nn\\
&=  G(p_1,p_2,p_3,p_4) \frac{1}{P_{1234}^2} G(p_5,p_6,p_7,p_8) + \ldots,
\label{Galileonamp}
\end{align}
where the ellipses denote permutations.  This expression is manifestly equal to the Feynman diagram expression.  

Note the similarity between the above manipulations and the derivation of the CSW rules for Yang-Mills amplitudes.  While MHV amplitudes are invariant under square bracket shifts, the Galileon vertices literally rescale under the rescaling shift.  Just as the CSW rules can be proven using the Risager three-line momentum shift \cite{Risager}, the Feynman diagram expansion of the general Galileon can be proven using our new recursion relations.

\section{Outlook}

We have derived a new class of recursion relations for effective field theories with enhanced soft limits, {\it i.e.}, the non-linear sigma model, Dirac-Born-Infeld theory, and the Galileon.   Like gauge and diffeomorphism invariance, soft behavior dictates non-trivial relations among interactions of different valencies.  

Our results open many avenues for future work \cite{Simplest}.  In particular, while we have considered fixed $\rho$ theories here, it should be straightforward to generalize our results to mixed $\rho$ theories such as the DBI-Galileon \cite{deRham}.  Also interesting would be to extend our results to theories with universal albeit non-vanishing soft behavior. For example, in the conformal Dirac-Born-Infeld model---corresponding to the motion of a brane in AdS---the soft limits of amplitudes are not zero but closely related lower-point amplitudes. Last but not least, there is the question of how recursion relations might utilize the behavior of amplitudes in other kinematical regions like collinear or double-soft limits (for recent discussion see~\cite{double}).
\smallskip

{\it Acknowledgment:} This work is supported in part by Czech Government project no.~LH14035 and GACR 15-18080S.
J.T.~is supported by the
David and Ellen Lee Postdoctoral Scholarship and DOE under grant no.~DE-SC0011632, and C.C.~and C.-H.S.~are supported by a Sloan Research Fellowship and a DOE Early Career Award under grant no.~DE-SC0010255. This work was performed in part at the Aspen Center for Physics, which is supported by National Science Foundation grant no.~PHY-1066293.


\begin{thebibliography}{99}

\bibitem{BCFW}

  R.~Britto, F.~Cachazo and B.~Feng,
  Nucl.\ Phys.\ B {\bf 715}, 499 (2005);
  R.~Britto, F.~Cachazo, B.~Feng and E.~Witten,
  Phys.\ Rev.\ Lett.\  {\bf 94} (2005) 181602

    \bibitem{GR} 
    J.~Bedford, A.~Brandhuber, B.~J.~Spence and G.~Travaglini,
  Nucl.\ Phys.\ B {\bf 721}, 98 (2005);
    F.~Cachazo and P.~Svrcek,
    hep-th/0502160.
    

\bibitem{Recursion}

  T.~Cohen, H.~Elvang and M.~Kiermaier,
  JHEP {\bf 1104} (2011) 053;
  C.~Cheung,
  JHEP {\bf 1003}, 098 (2010)


\bibitem{OurRec}

  C.~Cheung, C.~H.~Shen and J.~Trnka,
  JHEP {\bf 1506}, 118 (2015)
  
  	\bibitem{Grassmanian} 
  	N.~Arkani-Hamed, F.~Cachazo, C.~Cheung and J.~Kaplan,
  	JHEP {\bf 1003}, 020 (2010);
  	N.~Arkani-Hamed, J.~L.~Bourjaily, F.~Cachazo, A.~B.~Goncharov, A.~Postnikov and J.~Trnka,
  	arXiv:1212.5605 [hep-th].

  	\bibitem{amplitudhedron} 
  	N.~Arkani-Hamed and J.~Trnka,
  	JHEP {\bf 1410}, 30 (2014)
	\bibitem{pion} 
  J.~A.~Cronin,
  Phys.\ Rev.\  {\bf 161} (1967) 1483;
  S.~Weinberg,
  Phys.\ Rev.\ Lett.\  {\bf 18} (1967) 188;
  S.~Weinberg,
  Phys.\ Rev.\  {\bf 166} (1968) 1568.
	\bibitem{Cheung:2007st} 
	C.~Cheung, P.~Creminelli, A.~L.~Fitzpatrick, J.~Kaplan and L.~Senatore,
	JHEP {\bf 0803}, 014 (2008)
	
	\bibitem{OurEFT}
	
	C.~Cheung, K.~Kampf, J.~Novotny and J.~Trnka,
	Phys.\ Rev.\ Lett.\  {\bf 114}, no. 22, 221602 (2015)
	

\bibitem{Galileons}

G.~R.~Dvali, G.~Gabadadze and M.~Porrati,
Phys.\ Lett.\ B {\bf 485} (2000) 208;
A.~Nicolis, R.~Rattazzi and E.~Trincherini,
Phys.\ Rev.\ D {\bf 79} (2009) 064036;

	
	\bibitem{ElvangHuang} 
	H.~Elvang and Y.~t.~Huang,
	arXiv:1308.1697 [hep-th].
  	

	\bibitem{Boundary}
	B.~Feng, K.~Zhou, C.~Qiao and J.~Rao,
	JHEP {\bf 1503}, 023 (2015);
	  Q.~Jin and B.~Feng,
	  JHEP {\bf 1506} (2015) 018;
	  Q.~Jin and B.~Feng,
	  arXiv:1507.00463 [hep-th].

	\bibitem{NimaLargeZ}
	N.~Arkani-Hamed and J.~Kaplan,
	JHEP {\bf 0804}, 076 (2008);
	C.~Cheung,
	JHEP {\bf 1003}, 098 (2010)
	
\bibitem{NLSMlet}

  K.~Kampf, J.~Novotny and J.~Trnka,
  Phys.\ Rev.\ D {\bf 87}, no. 8, 081701 (2013)

  
  \bibitem{shiftsym}
  K.~Hinterbichler and A.~Joyce,
  Int.\ J.\ Mod.\ Phys.\ D {\bf 23}, no. 13, 1443001 (2014);
  T.~Griffin, K.~T.~Grosvenor, P.~Horava and Z.~Yan,
  arXiv:1412.1046 [hep-th].
  \bibitem{scatteq}
  F.~Cachazo, S.~He and E.~Y.~Yuan,
  arXiv:1412.3479.

  \bibitem{ambit}
  E.~Casali, Y.~Geyer, L.~Mason, R.~Monteiro and K.~A.~Roehrig,
  arXiv:1506.08771 [hep-th].


\bibitem{NLSMpap}

K.~Kampf, J.~Novotny and J.~Trnka,
JHEP {\bf 1305}, 032 (2013)



 \bibitem{hidden}

  K.~Hinterbichler and A.~Joyce,
  Phys.\ Rev.\ D {\bf 92}, no. 2, 023503 (2015);

\bibitem{Dualities}

  C.~De Rham, L.~Keltner and A.~J.~Tolley,
  Phys.\ Rev.\ D {\bf 90} (2014) 2,  024050; 
  P.~Creminelli, M.~Serone, G.~Trevisan and E.~Trincherini,
  JHEP {\bf 1502} (2015) 037.

\bibitem{DualitiesKN}  
K.~Kampf and J.~Novotny,
JHEP {\bf 1410} (2014) 006





\bibitem{Risager} 
  K.~Risager,
  JHEP {\bf 0512}, 003 (2005)
\bibitem{Simplest}
  C.~Cheung, K.~Kampf, J.~Novotny and J.~Trnka, {\it In prep.}
\bibitem{deRham} 
  C.~de Rham and A.~J.~Tolley,
  JCAP {\bf 1005}, 015 (2010)




\bibitem{double}
F.~Cachazo, S.~He and E.~Y.~Yuan,
arXiv:1503.04816 [hep-th];
T.~Klose, T.~McLoughlin, D.~Nandan, J.~Plefka and G.~Travaglini,
JHEP {\bf 1507} (2015) 135;
Y.~J.~Du and H.~Luo,
JHEP {\bf 1508}, 058 (2015)
P.~Di Vecchia, R.~Marotta and M.~Mojaza,
arXiv:1507.00938 [hep-th].


\end{thebibliography}
\end{document}